
\input phyzzx
\nopubblock
\line{January, 1995 \hfil hep-ph/9501388}
\line{\hfil BOW-PH-105}
\titlepage
\title{{\seventeenrm
Fermions Destabilize Electroweak Strings}}
\author{Stephen G. Naculich\footnote\star{e-mail:
naculich@polar.bowdoin.edu}}
\address{Department of Physics\break
         Bowdoin College\break
         Brunswick, ME  04011}
\abstract{ Z-strings in the Weinberg-Salam model
including fermions are unstable for all values of the parameters.
The cause of this instability is the fermion vacuum energy
in the Z-string background.  Z-strings with non-zero fermion
densities, however, may still be stable.  }
\endpage
\overfullrule=0pt
\def\PLB{ \sl Phys. Lett.         \bf B}
\def\NPB{ \sl Nucl. Phys.         \bf B}
\def\PRD{ \sl Phys. Rev.          \bf D}
\def\PRL{ \sl Phys. Rev. Lett.    \bf  }
\def\d { {\rm d} }
\def\e { {\rm e} }
\def\ppm{ \psi_\pm }
\def\P{ P_\pm}
\def\Ee{E_{\rm effective}}
\def\Eb{E_{\rm boson}}
\def\Ef{E_{\rm fermion}}
\def\Efz{E_{{\rm fermion}}^\prime}
\def\Cb{C_{\rm b}}
\def\Cf{C_{\rm f}}
\def\eps{\epsilon}
\def\del{\partial}
\def\Dslash{\rlap{D}\,/}
\def\Phitil{{\tilde \Phi}}
\def\Psibar{{\overline \Psi}}
\def\psibar{{\overline \psi}}
\def\rrho{\rho^\prime}

\REF\rrNambu{Y. Nambu, \NPB 130 \rm (1977) 505.}
\REF\rrVach{T. Vachaspati, \PRL 68 \rm (1992) 1977, \bf 69
		\rm (1992) 216(E).}
\REF\rrHHVW{R. Holman, S. Hsu, T. Vachaspati, and R. Watkins,
		\PRD 46 \rm (1992) 5352.}
\REF\rrBaryo{R. Brandenberger and A. Davis, \PLB 308 \rm (1993) 79;\nextline
M. Barriola, \PRD 51 \rm (1995) 300.}
\REF\rrJPV{M. James, L. Perivolaropoulos, and T. Vachaspati,
		\PRD 46 \rm (1992) 5232; \NPB 395 \rm (1993) 534.}
The recent discovery of cosmic string solutions
in the Weinberg-Salam model of electroweak interactions
[\rrNambu, \rrVach]
has fueled a burst of activity in the study of defects
in the standard model and its extensions,
and of their possible consequences
for astrophysics and cosmology [\rrHHVW, \rrBaryo].
The existence of these ``electroweak strings,''
which are essentially Nielsen-Olesen vortices
embedded in the Weinberg-Salam model,
was previously neglected
because they do not possess topological stability.
They may nonetheless be stable
if they sit at a local minimum of the energy.
Because they owe their existence to energetic
rather than topological reasons,
the stability of electroweak strings
is sensitively dependent on the field content
and the values of the parameters in the theory.
For example,
in a simplified version of the Weinberg-Salam model
containing only bosonic fields,
Z-strings [\rrVach] are stable
only for light Higgs masses ($\le m_Z$),
and for $\sin^2 \theta_W$ fairly close to unity [\rrJPV],
a region that obviously does not include the physical world.

\REF\rrVW{T. Vachaspati and R. Watkins, \PLB 318 \rm (1993) 163.}
\REF\rrStable{M. Earnshaw and M. James, \PRD 48 \rm (1993) 5818;\nextline
L. Perivolaropoulos, \PLB 316 \rm (1993) 528;\nextline
G. Dvali and G. Senjanovi\'c, \PRL 71 \rm (1993) 2376.}
Attempts have been made to increase the range of stability
of the Z-string by extending the model to include other fields
[\rrHHVW, \rrVW, \rrStable].
One idea, familiar from the study of nontopological solitons,
is to include particles whose mass arises from the Higgs mechanism.
Such particles remain massless at the center of the string
where the Higgs field vanishes,
and the presence of such particles at the core
would resist the string's dissolution,
because that would increase their energy.
Indeed, the presence of charged
scalar bound states was shown
to lower the value of $\sin^2 \theta_W$
for which the string is stable [\rrVW].

\REF\rrEP{M. Earnshaw and W. Perkins, \PLB 328 \rm (1994) 337.}
\REF\rrGV{J. Garriga and T. Vachaspati, \NPB 438 \rm (1995) 161.}
\REF\rrMOQ{J. Moreno, D. Oaknin, and M. Quir\'os, \PLB  347 \rm (1995) 332.}
It has been suggested that
a similar enhancement of stability
could be achieved by using fermion bound states on the Z-string
[\rrVW].
The existence of Z-string zero modes,
fermion states localized on the string with zero energy,
lends support to this idea [\rrEP--\rrMOQ].
Another advantage of this suggestion is
that fermions are already contained in
the standard electroweak model!

In this letter,
we show that, on the contrary,
the presence of fermions in the electroweak theory
{\it destabilizes} Z-strings.
More precisely,
the lowest-energy (or ground) state of the Z-string
is always a local maximum of the energy functional
with respect to (at least) one of the modes of instability.
The Z-string is therefore unstable
for {\it all} values of the parameters
of the Weinberg-Salam model.
(It is possible, however,
that a higher-energy state of the Z-string,
with a finite quark density,
could be locally stable.)

\REF\rrVacuum{J. Bagger and S. Naculich, \PRL 67 \rm (1991) 2252;
	\PRD 45 \rm (1992) 1395;\nextline
	S. Naculich, \PRD 46 \rm (1992) 5487.}
This instability results from the fermion vacuum energy,
which also has an important effect
on other types of solitons [\rrVacuum].
One cannot consistently consider
the effects of positive-energy fermion states
without also taking account of the
(filled) negative-energy states,
particularly because,
with the existence of zero modes,
there is no gap between them.
We will show that the contribution
to the energy functional of the filled Dirac sea,
\ie, the fermion vacuum energy,
is a local maximum for the Z-string.

First, we will describe the fermion spectrum
in the presence of the Z-string;
then we will show how the fermion vacuum energy
changes under certain small perturbations
away from the Z-string.
The electroweak Lagrangian is
$$
L = L_{\rm boson} + \sum L_{\rm quark} + \sum L_{\rm lepton}
\eqn\eeLag
$$
with
$$
L_{\rm boson} =
  -{1 \over 4} W_{\mu \nu}^a W^{a \mu \nu}
  -{1 \over 4} F_{\mu \nu}   F^{\mu \nu}
  + \left| D^L_\mu  \Phi \right|^2
  - \lambda \left( \Phi^\dagger \Phi - {\eta^2 \over 2} \right)^2
\eqn\eeLboson
$$
where $W^a_{\mu \nu}$ and $F_{\mu \nu}$
are field strength tensors for
the SU(2)$_L$ and U(1)$_Y$ gauge fields
$W^a_\mu$ and $B_\mu$ respectively,
and
$\Phi  = \left( \phi_1 \atop \phi_0 \right) $
is the complex Higgs doublet.
Each quark doublet contributes a term
$$
\eqalign{
L_{\rm quark}  =
& \Psibar^L i \Dslash^L \Psi^L
+ \psibar_+^R i \Dslash^R \psi_+^R
+ \psibar_-^R i \Dslash^R \psi_-^R \cr
& - {G_+} \left( \Psibar^L \Phitil \psi_+^R
            + \psibar_+^R \Phitil^\dagger \Psi^L \right)
- {G_-}  \left( \Psibar^L \Phi \psi_-^R
            + \psibar_-^R \Phi^\dagger \Psi^L \right) \cr}
\eqn\eeLquark
$$
where $\Psi^L = \left( \psi_+^L \atop \psi_-^L \right)$,
and
$\Phitil = i \tau^2 \Phi^* = \left( \phi_0^* \atop -\phi_1^* \right) $.
Each lepton doublet contributes the same term,
absent any pieces containing $\psi_+^R$.
(We neglect interfamily mixing.)
The gauge-covariant derivatives in eqs. \eeLag~and \eeLboson~are
$$
D_\mu^L
= \del_\mu - {ig\over 2} \tau^a W^a_\mu  - {i g^\prime \over 2} Y B_\mu,
\qquad
D_\mu^R
= \del_\mu - {i g^\prime \over 2} Y B_\mu,
\eqn\eeDeriv
$$
with $Y$ the hypercharge of the field on which the
covariant derivative is acting.

\REF\rrNO{H. Nielsen and P. Olesen, \NPB 61 \rm (1973) 45.}
The Z-string [\rrVach] is the field configuration
$$
\Phi = {\eta f(\rho) \over \sqrt 2} \e^{i\phi} \left( 0 \atop 1 \right),\qquad
\left( Z^1 \atop Z^2 \right)
 = {2 v(\rho) \over \alpha \rho} \left( -\sin \phi \atop \cos \phi \right),
\eqn\eeZstring
$$
all other fields vanishing,
where $f(\rho)$ and $v(\rho)$
obey the Nielsen-Olesen equations [\rrNO]
$$
\eqalign{
 f^{\prime\prime}
+ {f^\prime \over \rho}
- (1-v)^2 {f \over \rho^2}
+ \lambda \eta^2 (1-f^2) f & = 0, \cr
 v^{\prime\prime}
 - {v^\prime \over \rho}
+ {\alpha^2 \eta^2 \over 4} f^2 (1-v) & = 0}
\eqn\eeNO
$$
with boundary conditions
$$
f(0)=  v(0) = 0,
\qquad
f(\rho) \mathrel{\mathop{\longrightarrow}\limits_{\rho \to \infty}} 1,
\qquad
v(\rho) \mathrel{\mathop{\longrightarrow}\limits_{\rho \to \infty}} 1.
\eqn\eeBoundary
$$
Recall that
$Z_\mu = \cos \theta_W \, W^3_\mu - \sin \theta_W \, B_\mu$
and
$\alpha = \sqrt{g^2 + {g^\prime}^2}$.

The main question is whether such a field configuration
is energetically stable.
One possible mode of instability
is that the upper component $\phi_1$ of the Higgs field
may develop a non-zero value,
allowing the Z-string to unwind.
(There may be other modes of instability as well.)
To determine whether the string is stable
to small perturbations in this direction
(in the absence of fermions),
one computes the change in the bosonic field energy
$$
\Delta \Eb [\phi_1] = \Eb [f, v; \phi_1] - \Eb [f, v; 0].
\eqn\eeDeltaEb
$$
If the Z-string is stable,
$\Eb [f,v;0]$ is a local minimum of the energy functional,
and $\Delta \Eb [\phi_1]$ will be quadratic in $\phi_1$
with a positive coefficient [\rrVach].

\REF\rrJRos{R. Jackiw and P. Rossi, \NPB 190 [FS3] \rm (1981) 681.}
Next we consider the effect of fermions in the electroweak theory.
The Dirac equation in the Z-string background has the form
$$
\eqalign{
& \gamma^\mu ( i \del_\mu - {\alpha \ell_{\pm}\over 2} Z_\mu ) \ppm^L
- m_\pm f(\rho) \e^{\mp i\phi} \ppm^R = 0, \cr
&\gamma^\mu ( i \del_\mu - {\alpha r_{\pm}\over 2} Z_\mu ) \ppm^R
- m_\pm f(\rho) \e^{\pm i\phi} \ppm^L = 0, \cr}
\eqn\eeDirac
$$
where $m_\pm = G_\pm \eta / \sqrt2$,
$\ell_\pm = (y \pm 1) \sin^2 \theta_W \mp 1 $,
and $   r_\pm = (y \pm 1) \sin^2 \theta_W $,
with $y$ the hypercharge of the left-handed doublet $\Psi^L$.
This equation has zero-energy modes [\rrJRos, \rrEP--\rrMOQ],
which obey $ \gamma_0 \gamma_3 \ppm = \pm \ppm $.
Using the chiral representation for the Dirac matrices
$$
\gamma^5 = \pmatrix{ 1 & 0  \cr 0 & -1 \cr}, \quad
\gamma^0 = \pmatrix{ 0 & -1 \cr -1 & 0 \cr}, \quad
\gamma^i = \pmatrix{ 0 & \tau^i \cr -\tau^i & 0 \cr},
\eqn\eeGamma
$$
and recalling that
$\psi_\pm^L = {1\over 2} (1 -\gamma^5) \psi_\pm$
and
$\psi_\pm^R = {1\over 2} (1 +\gamma^5) \psi_\pm$,
one may write the zero mode solutions as
$$
\psi_{\pm,0}  ~=~
\e^{ \pm \int_0^\rho  \left[\ell_\pm v(\rrho) / \rrho \right] \d \rrho }
\left( {i \over m_\pm f(\rho)} \P^\prime(\rho) \chi_\pm \atop
        \P (\rho) \chi_\mp \right)
\eqn\eeZeromode
$$
where $\chi_+ = \left( 1 \atop 0 \right)$, $\chi_- = \left( 0 \atop 1 \right)$,
with $\P(\rho)$ obeying the equation
$$
\P^{\prime\prime}
\, - \, {f^\prime \over f} \P^\prime
\, \pm \, {(\ell_\pm + r_\pm) v\over \rho } \P^\prime
\, - \,  m^2_\pm f^2 \P \, = \, 0
\eqn\eeP
$$
and normalized by
$$
\int \d^3 x \,
\e^{ \pm 2 \int_0^\rho  \left[\ell_\pm v(\rrho) / \rrho \right] \d \rrho }
\left[ \P^2 + \left( \P^\prime/ m_\pm f \right)^2 \right] = 1
\eqn\eeNorm
$$
For the neutrino ($m_+ = 0$, $\ell_+ = -1$),
eq.~\eeP~has the simple solution $P_+ = 1$,
but by eq.~\eeBoundary,
$ \psi_{+,0}
\mathrel{\mathop{\longrightarrow}\limits_{\rho \to \infty}}
1/\rho$,
so the zero mode is not normalizable,
at least for a straight infinite string
(but see ref.~[\rrGV]).
Eq.~\eeP~has the explicit solution
$$
\P (\rho) = N \e^{ - \int_0^\rho  m_\pm f(\rrho) \d \rrho }
\eqn\eeExplicit
$$
for the special case $ y=0$ and $\cos^2 \theta_W = {1\over2}$.

\REF\rrJReb{R. Jackiw and C. Rebbi, \PRD 13 \rm (1976) 3398.}
The existence of zero modes generates a
$2^N$-fold degeneracy of the Z-string ground state,
where $N$ is the number of quark and charged lepton flavors.
The ground state of the string
will have the global quantum numbers of each fermion flavor [\rrJReb],
either $\half$ or $-\half$,
depending on whether the corresponding zero mode is occupied or not.
The occupation of the zero modes will
not alter the Nielsen-Olesen equations \eeNO~for the string profile,
because the fermion source term for the $\phi_0$ and $Z_\phi$ fields
vanishes for the zero modes.

In the $3+1$-dimensional context of the Z-string,
the zero-energy solution \eeZeromode~generates
a whole family of solutions of the Dirac equation
$$
\psi_{\pm,p} (\rho,z,t)
{}~=~  \e^{ ipz - i\eps_{\pm,p} t } ~\psi_{\pm,0} (\rho)
\eqn\eeMassless
$$
with energies
$$
\eps_{\pm, p} = \pm p .
\eqn\eeDispersion
$$
These solutions correspond to
massless chiral fermions confined to the Z-string;
the up-type quarks run up the string (in the $+z$ direction)
and the down-type quarks and charged leptons run down the string
(in the $-z$ direction) at the speed of light.
In addition to these ``massless'' solutions of the Dirac equation,
there are many ``massive'' solutions,
whose energies are separated from zero by a finite gap.

What effect have fermions on the stability of the Z-string?
Earnshaw and Perkins [\rrEP]
pointed out that the fermion zero mode
provides a non-vanishing source term
in the equation of motion for $\phi_1$.
This violates the ``Vachaspati existence criterion'' [\rrVach]
and would appear to imply that
the Z-string configuration with $\phi_1=0$
is not an extremum of the energy.
Such a conclusion, however, would be premature.

The reason that the zero modes are a source for $\phi_1$
is that the presence of a non-zero value of $\phi_1$
lifts the degeneracy between the
$\psi_{+,0}$ and $\psi_{-,0}$ zero modes,
one linear combination of the zero modes shifting up
and the orthogonal combination shifting down.
The lower eigenstate is filled in the ground state of the Z-string,
so its descent lowers the Z-string energy.
Before drawing any conclusions about the overall stability
of the Z-string, however,
we must determine the effect of the $\phi_1$ perturbation
on the rest of the fermion eigenenergies.

The effective energy of the Z-string ground state is
$$
\Ee = \Eb + \Ef
\eqn\eeEffective
$$
where $\Eb$ is the bosonic field energy
and $\Ef$ the fermion vacuum energy in the Z-string background
(\ie, the energy of the filled Dirac sea).
The change $\Delta \Eb [\phi_1]$
under a small perturbation $\phi_1$
was considered above \eeDeltaEb;
the change in the fermion vacuum energy is
$$
\Delta \Ef [\phi_1]
= \sum_{ \eps_{+,n} <0}  \Delta \eps_{+,n} [\phi_1]
+ \sum_{ \eps_{-,n} <0}  \Delta \eps_{-,n} [\phi_1]
+ \delta E [\phi_1]
\eqn\eeDeltaEf
$$
where
$ \Delta \eps_{\pm,n} [\phi_1] $
denotes the shift of the Z-string Dirac eigenenergies
$\eps_{\pm, n}$ under the perturbation,
and the sum is over negative-energy eigenvalues only.

We compute the eigenvalue shifts
$ \Delta \eps_{\pm,n} [\phi_1] $
perturbatively in $\phi_1$.
Because the perturbation is off-diagonal in the $+$ and $-$ fields,
the leading shift is second order,
and the change in fermion vacuum energy is
$$
\eqalign{
\Delta \Ef [\phi_1]
&=
 \sum_{\eps_{+,n}<0} \sum_{\eps_{-,m}>0}
 {\left|  \int \d^3 x
 \left( {G_-} \psibar_{-,m}^R  \psi_{+,n}^L
- {G_+} \psibar_{-,m}^L  \psi_{+,n}^R  \right)
\phi_1^* \right|^2 \over \eps_{+,n} - \eps_{-,m} }
\cr
& +
\sum_{\eps_{-,n}<0} \sum_{\eps_{+,m}>0}
{\left|  \int \d^3 x
  \left( {G_-} \psibar_{+,m}^L  \psi_{-,n}^R
       - {G_+} \psibar_{+,m}^R  \psi_{-,n}^L \right)
 \phi_1 \right|^2 \over \eps_{-,n} - \eps_{+,m} } + \delta E [\phi_1].
\cr
}
\eqn\eeSecond
$$
The sums over intermediate energy eigenstates $\eps_{\pm, m}$
include only positive-energy states;
the contributions from negative-energy intermediate states
cancel between the two sums.
The sums in eq.~\eeSecond~diverge in the ultraviolet.
The Z-string is not responsible for this,
for the same divergence occurs
in the usual constant field background.
In that case,
the divergence is cancelled
by adding a counterterm $\delta E [\phi_1]$.
The same counterterm will suffice to
render $\Delta \Ef [\phi_1]$ ultraviolet finite.

Let us evaluate the shifts in eigenenergies
corresponding to the massless solutions \eeMassless~more explicitly.
First, restrict the perturbation $\phi_1$
to a constant (complex) value $\eta g /\sqrt2 $ over the region
where the zero mode wavefunction \eeZeromode~is appreciable
(but let $\phi_1 \to 0$ as $\rho \to \infty$).
Second,
noting that $\Ef$
is proportional to the length of the string (as is $\Eb$),
consider a Z-string of length $L$.
Periodic boundary conditions on the fermion wavefunctions
restrict the $z$-momenta to $ p = 2\pi n / L$.
Taking $L$ large,
the sum of the energy shifts of the massless states becomes
$$
\Delta \Efz [g]
 = -{L \over \pi} |g|^2 |A|^2 \int_0^\Lambda {\d p \over 2p}
\eqn\eeIRdivergent
$$
with
$$
A= 2\pi iL \int \rho \d \rho
{ \left( P_+ P_- \right)^\prime \over f}
\exp \left({\int_0^\rho \left[ (\ell_+ - \ell_-) v(\rrho)
               / \rrho \right] \d \rrho}\right)
\eqn\eeA
$$
where we have included only massless intermediate energy states.
Among this subset of intermediate states,
a selection rule ensures that the perturbation
only couples the eigenstate $\psi_{\pm,p}$
to the eigenstate $\psi_{\mp,-p}$.

\REF\rrCY{S.-J. Chang and T.-M. Yan, \PRD 12 \rm (1975) 3225.}
The integral over momenta \eeIRdivergent~diverges
both in the ultraviolet and in the infrared.
As previously noted,
the ultraviolet divergence is cancelled by counterterms;
the infrared divergence signals
the breakdown of the perturbative expansion
when the energy denominator $2p$
becomes smaller than the perturbation $g$.
We redo the calculation for states with small $p$,
now treating $p$ as part of the perturbation.
The unperturbed states are now degenerate;
degenerate perturbation theory yields the perturbed energies
$$
\left| \matrix{
\eps - p  & g A   \cr  g^*  A^*  & \eps + p \cr} \right| = 0
\qquad \Rightarrow \qquad
\eps = \pm \sqrt{ p^2 + |gA|^2 }
\eqn\eeDegenerate
$$
As mentioned above,
the degenerate zero mode ($p=0$) states
are resolved into $ \eps = \pm |gA|$.
This improved calculation yields an infrared-finite result
$$
\eqalign{
\Delta \Efz [g]
&=
- {L\over 2\pi} \int_{-\Lambda}^\Lambda \d p
\left( \sqrt{ p^2 + |gA|^2 } - |p| \right)
\cr
& \mathrel{\mathop{\longrightarrow}\limits_{\Lambda \to \infty}}
- {L \over 4 \pi} |gA|^2
\left[1+\log \left( 4\Lambda^2 \over |gA|^2 \right) \right]
\cr}
\eqn\eeIRfinite
$$
The ultraviolet divergence is absorbed by the counterterm,
leading to a completely finite expression
for the change in the fermion vacuum energy
(per unit length) under the perturbation $\phi_1$:
$$
\Delta \Ef [g]
=  L \left(  {|A|^2 \over 4 \pi} |g|^2  \log |g|^2 + \Cf |g|^2 \right)
\eqn\eeDeltaEfg
$$
The coefficient $\Cf$ receives contributions
from the shifts of the massive Dirac eigenvalues
as well as from the finite part of the counterterm.
Each quark doublet contributes a term of the form \eeDeltaEfg~to
the fermion vacuum energy (with different values of $A$);
the charged leptons do not contribute
because their eigenenergies are not shifted by the perturbation
(in the absence of normalizable neutrino zero modes).

The change in the bosonic field energy
for the perturbation we are considering
has the form [\rrVach]
$$
\Delta \Eb [g] = L \Cb |g|^2
\eqn\eeDeltaEbg
$$
where the sign of $\Cb$,
which depends on the parameters of the model,
determines whether the bosonic Z-string (without fermions)
is stable in the direction of this perturbation.
Thus, the effective energy of the ground state of the
Z-string is
$$
\Ee [g]
= \Ee [0]
+  L \left(  C |g|^2
+ {1 \over 4 \pi}
|g|^2 \log |g|^2
\sum_{\rm quark} |A|^2
\right)
\eqn\eeEffectiveg
$$
Observe first that
$|g| = 0$
is an extremum of this expression,
so even in the presence of fermions
the Z-string configuration \eeZstring~with $|\phi_1| = 0$
remains a solution of the equations of motion.
This extremum, however,
is necessarily a maximum,
regardless of the value of $C$.
(See ref.~[\rrCY] for a similar phenomenon in two dimensions.)
Hence, the Z-string is unstable
to perturbations in $\phi_1$
for {\it all} values of the parameters of the Weinberg-Salam model.

For a Z-string of large but finite length $L$,
the fermion energy \eeIRfinite~becomes a sum over momenta,
given by
$$
\Delta \Efz [g]
= - | gA |
 - { L |gA|^2  \over 2 \pi }
\left[ \log \left(L \Lambda \over 2\pi\right) + \gamma \right] + \cdots
$$
in the limit $ g \ll 1/L$.
The leading term linear in $g$ can be cancelled
by populating both (or neither) zero modes.
The subleading term then contributes energy per unit length
quadratic in $g$ with a negative coefficient
that diverges as $\log L$.
Thus, regardless of the bosonic contribution \eeDeltaEbg,
the Z-string ground state is unstable for large $L$,
the same conclusion reached above.

This does not mean that
there exists no stable nontopological string configuration.
A (locally) stable string with $\phi_1$ slightly displaced
from zero may exist, though that remains to be demonstrated.
What we are saying is that
the simple Nielsen-Olesen string embedded
into the Weinberg-Salam model
with all other fields vanishing
is necessarily unstable.

Most attempts to increase the stability of the Z-string
do so by increasing the coefficient $C$ in \eeEffectiveg;
this will not work here since no coefficient,
however positive,
can outweigh the negative curvature at $|g|=0$
caused by the $|g|^2 \log |g| $ term.
The only way to overcome this term
is to occupy some of the {\it positive}-energy fermion states.
This requires not just a single fermion
(as in the case of nontopological solitons)
but rather a finite density of fermions along the Z-string.
If the string holds $\zeta$ positive-energy fermions
per unit length of types $+$ and $-$,
the effective energy will change by
$$
\eqalign{
\Delta E [g]  &
=
 {L\over 2\pi} \int_{-2\pi\zeta}^{2\pi\zeta} \d p
\left( \sqrt{ p^2 + |gA|^2 } - |p| \right)
\cr
&
\mathrel{\mathop{\longrightarrow}\limits_{ |gA| \ll 2\pi\zeta}}
 {L \over 4 \pi} |gA|^2
\left[1+\log \left( 16 \pi^2 \zeta^2 \over |gA|^2 \right) \right]
\cr}
\eqn\eeFinitedensity
$$
If the Z-string carries
a non-zero density of quarks of {\it each} flavor,
the change in energy
will cancel the $ |g|^2  \log |g|^2 $ piece in \eeEffectiveg,
rendering the total energy proportional to $|g|^2$.
A difficult calculation would then be required
to determine the minimum fermion density
necessary to stabilize the state.
Even so, the state might only be a local minimum,
liable to decay to a non-string state via
tunnelling or thermal fluctuations.

\centerline{\fourteenrm Acknowlegments}
It is a pleasure to thank T. Vachaspati for
several useful discussions.

\endpage
\refout
\end